\documentclass[twocolumn, aps,amsmath, amssymb, superscriptaddress]{revtex4}

\usepackage{dcolumn}
\usepackage{bm}
\usepackage{amsmath}
\usepackage{amsfonts}
\usepackage{graphics}
\usepackage[dvips]{graphicx}
\usepackage{times}

\begin{document}

\title{Quantitative and empirical demonstration of the Matthew effect in a study of career longevity}

\author{Alexander M. Petersen}
\affiliation{Center for Polymer Studies and Department of Physics, Boston University, Boston, Massachusetts 02215, USA}
\author{Woo-Sung Jung}
\affiliation{Graduate Program for Technology and Innovation Management and Department
of Physics, Pohang University of Science and Technology, Pohang 790-784,
Republic of Korea}
\affiliation{Center for Polymer Studies and Department of Physics, Boston University, Boston, Massachusetts 02215, USA}
\author{Jae-Suk Yang}
\affiliation{Sanford C. Benstein \& Co. Center for Leadership and Ethics, Columbia Business School, Columbia University, New York, NY 10027}
\author{H. Eugene Stanley}
\affiliation{Center for Polymer Studies and Department of Physics, Boston University, Boston, Massachusetts 02215, USA}

\date{\today}

\begin{abstract}
The Matthew effect refers to the adage written some two-thousand years ago in the Gospel of St. Matthew: 
``For to all those who have, more will be given."
 Even two millennia later, this idiom is used by sociologists to qualitatively describe the dynamics of individual progress and the interplay between status and reward. Quantitative studies of
professional careers are traditionally limited by the difficulty in measuring progress and the lack of data on
individual careers. However, in some professions, there are well-defined metrics that quantify career longevity,
success, and prowess, 
 which together contribute to the overall success rating for an individual employee.  
Here we demonstrate  testable evidence of the age-old Matthew ``{\it rich get richer}"  effect, wherein the longevity and past success of an individual lead to a
cumulative advantage in further developing his/her career.   
We develop an exactly solvable  stochastic career progress model  that quantitatively incorporates  the Matthew  effect, and validate our model predictions for several competitive professions.  We test our model
on the careers of $400,000$ scientists using data from six high-impact journals, and further confirm our findings by
testing the model on the careers of more than $20,000$ athletes in four sports leagues. 
Our model highlights the importance of early career development, showing that many careers are stunted by the relative disadvantage associated with inexperience.
\end{abstract}

\maketitle 
\footnotetext[1]{ Corresponding Author: Alexander M. Petersen \newline
\emph{E-mail}: amp17@physics.bu.edu}

The rate of individual progress is fundamental to career development and  success. In practice, the rate of
progress depends on many factors, such as an individual's talent, productivity, reputation as well as other external random factors. 
Using a stochastic model, here we find that the relatively small rate of progress at the beginning of the career plays a
crucial role in the evolution of the career length.  Our quantitative model
describes career progression using two fundamental ingredients: (i) random
forward progress ``up the career ladder", and  (ii) random stopping times, terminating a career. This model quantifies
the ``Matthew effect" by incorporating into ingredient (i) the  common cumulative advantage property \cite{Matthew,AccumAdvProdDiff,DeSollaPrice,CumAdvInequality,MatthewIII,MatthewEducation,MatthewReading,MatthewCountries} that it is easier to move forward in the career the further
along is one in the career. A direct result of the increasing progress rate with career position
is  the  large disparity between the numbers of  careers that are successful long tenures and the numbers of careers that are unsuccessful short stints.
 
 Surprisingly, despite the large differences in the numbers of long and short careers, we find a scaling law which bridges the gap between the frequent short and the infrequent long careers.
 We test this model for both scientific and sports careers,  two careers where accomplishments  are methodically
recorded. We analyze publication careers within six high-impact journals: Nature, Science, the Proceedings of the National Academy
of Science (PNAS),  
Physical Review Letters (PRL), New England Journal of Medicine (NEJM) and CELL. We also analyze sports careers within
four distinct leagues: Major League Baseball (MLB), Korean Professional Baseball, 
the National Basketball Association (NBA),  and the English Premier League.

 Career longevity is a fundamental metric that influences the overall legacy of an employee because for
most individuals the measure of success is intrinsically related, although not perfectly correlated, to his/her career
length. Common experience in most professions indicates that time is required
for colleagues to gain faith in a newcomer's abilities. Qualitatively, the acquisition of new opportunities mimics a
standard positive feedback mechanism (known in various fields as Malthusian growth, cumulative advantage, preferential
attachment, a reinforcement process, the ratchet effect, and the Matthew {\it ``rich get richer"} effect \cite{MatthewII}),
which endows greater rewards \cite{ColeCole} to individuals who are more accomplished than to individuals who are less accomplished. 

Here we use career position as a proxy for individual accomplishment, so that the positive feedback
captured by the Matthew effect is related to increasing career position.
There are also other factors that result in selective bias, such as  the ``relative age
effect", which has been used to explain
the skewed birthday distributions in populations of athletes. Several studies find that being born in optimal months
provides a competitive advantage to the older group members with respect to the younger group members within a cohort, resulting in a relatively higher chance of succeeding for the older group members, consistent with the Matthew effect. This relative age effect is found at 
several levels of competitive sports ranging from secondary school to the professional level \cite{RAFsoccer,RAFsport}.

In this paper we study the everyday topic of career longevity, and reveal surprising complexity arising from the generic competition
within social environments. We develop an exactly solvable stochastic model, which predicts the functional form  of the
probability density function (pdf)  $P(x)$ of career longevity $x$ in competitive professions, where we define career longevity as the final career position $x$  after a given time
duration $T$ corresponding to the termination time of the career. Our stochastic model depends on only two parameters, $\alpha$ and $x_{c}$. 
The first parameter, $\alpha$, represents the power-law exponent that emerges from the pdf of career longevity. This
parameter is intrinsically related to the progress rate  early in the career during which professionals establish their
reputations and secure future opportunities. The second parameter, $x_{c}$, is an effective time scale which distinguishes
newcomers from veterans. 


\begin{figure}[t]
\centerline{\includegraphics*[width=0.45\textwidth]{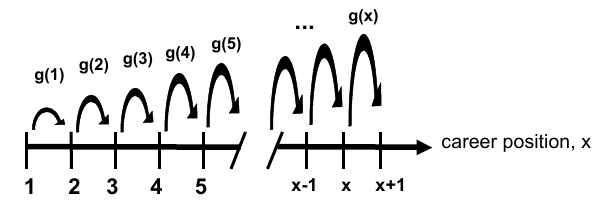}}
\caption{  Graphical illustration  of the stochastic Poisson process quantifying career progress with
position-dependent
progress rate $g(x)$ and stagnancy rate $1-g(x)$.  A new opportunity, corresponding to the advancement to career position $x+1$ from career position $x$, can refer to a day at work or, even more
generally, to
any assignment given by an employing body. In this framework, career progress is made at a rate $g(x)$ that is slower
than the passing of work-time, representing the possibility of career stagnancy.
The traditional Poisson process corresponds to a constant progress rate $g(x) \equiv \lambda$. Here, we use a functional
form for $g(x) \equiv 1-\exp[-(x/x_{c})^{\alpha}]$ that is increasing with career position $x$, which captures the
salient feature of the Matthew effect, that it becomes easier to make progress the further along the career.  In the Supporting Information text we further develop an alternative model where the progress rate $g(t)$ depends on time.
\label{fig0} } 
\end{figure} 

\section{Quantitative Model}

In this model, every employee begins his/her career with approximately zero credibility, and must labor through a common
development curve. 
At each position $x$ in a career, there is an opportunity for progress as well as the possibility for no progress. 
 A new opportunity, corresponding to the advancement to career position $x+1$ from career position $x$, can refer to a day at work or, more
generally, to
any assignment given by an employing body. For each particular career, the change in career position $\Delta x$ has an
associated time-frame $\Delta t$. Optimally, an individual makes progress by advancing in career position at an equal rate as the
 advancing of time $t$ so that $\Delta x \equiv
\Delta t$. However, in practice, an individual makes progress $\Delta x$ in a
subordinate time
frame, given here as the career position $x$. In this framework, career progress is made at a rate that is slower than
the passing of work time, representing the possibility of career stagnancy. 

As a
first step, we postulate that the stochastic process governing career progress is similar to a Poisson process, where
progress
is made at any given step with some approximate probability or rate. Each step forward in career position contributes to the employee's resume and
reputation. Hence, we refine the process to a spatial Poisson process, where
the probability of progress $g(x)$
depends explicitly on the employee's position $x$ within the career. 
 In our model, the progress rate $g(x) = talent(x) + reputation(x)+ productivity(x)+...$ represents a combination of several factors, such as the talent, reputation, and productivity at a given career position $x$. The criteria for  the Matthew effect to apply is that the  progress rate be monotonically increasing  with career position, so that $g(x+1) > g(x)$. In this paper, we do not distinguish between the Matthew effect, relating mainly to the positive feedback from recognition,  and cumulative advantage, which relates to the positive feedback from both productivity and recognition \cite{AccumAdvProdDiff}. 
 It would require more detailed data to determine the role of the individual factors on the evolution of a career.  
 
Employees begin their career at the starting career position $x_{0}\equiv 1$, and make random forward progress through
time to career position $x \geq 1$, as illustrated in Fig.~1.
Career longevity is then defined as the final
location $x \equiv x_{T}$ along the career ladder at the time of retirement $T$.
Let $P(x\vert T)$ be the conditional probability that at stopping time $T$ an individual is at  the final career position $x_{T}$. 
For simplicity, we assume that the progress rate $g(x)$ depends only on $x$.   As a result,  $P(x\vert T)$ assumes the
familiar Poisson form, but with the insertion of $g(x)$ as the rate parameter,
\begin{eqnarray}
 P(x\vert T) = \frac{e^{-g(x)T}(g(x)T)^{x-1}}{(x-1)!} \ .
 \label{Pxt}
\end{eqnarray} 
 We derive the 
spatial Poisson pdf  $P(x\vert T)$ in the {\it Appendix}. 
 In the Supporting Information (SI) text we further develop an alternative model where the progress rate  $g(t)$ represents a career trajectory which depends on time \cite{CareerTrajectory}.

According to the Matthew effect, it becomes  easier for an individual to excel with increasing success and reputation.
Hence, the choice of $g(x)$ should reflect the fact that newcomers, lacking the familiarity of their peers, have a more
difficult time moving forward,  
while seasoned veterans,  following from their experience and reputation, often have an easier time moving forward. For
this reason we choose
the progress rate $g(x)$ to have the functional form, 
\begin{eqnarray}
 g(x) \equiv  1- \exp[-(x/x_{c})^\alpha] \ . 
 \label{gx}
\end{eqnarray} 
This function exhibits the fundamental feature of increasing from approximately zero and asymptotically approaching
unity over some time interval $x_{c}$. Furthermore, $g(x) \sim x^{\alpha}$  for small $x \ll x_{c}$.
 In Fig.~2, we plot $g(x)$ for several values of $\alpha$, with fixed $x_{c} = 10^{3}$ in arbitrary units. 
We will show that the parameter $\alpha$ is the same as the power-law exponent $\alpha$ in the pdf of career longevity
$P(x)$, which we plot in the Fig.~2 inset. 
The random process  for forward progress can also be recast into the form of random waiting times, where the average
waiting time $\langle \tau(x)
\rangle$ between successive steps is the inverse of the forward progress probability, $\langle \tau(x) \rangle =
1/g(x)$.

We now address the fact that not every career is of the same length. 
Nearly every individual is faced with the constant risk of losing his/her job, possibly as the result of poor
performance, bad health, economic downturn, or even a change in the business strategy of his/her employer. 
Survival in the workplace requires that the individual
 maintain his/her performance level with respect to all possible replacements. In general, career longevity is
influenced by many competing random processes which contribute to the random termination time $T$ of a career
\cite{firing}. Our model accounts for external termination factors which are not correlated to the contemporaneous productivity of a given individual. A more sophisticated model, which incorporates endogenous termination  factors, e.g. termination due to sudden decrease in productivity below a given employment threshold, is more difficult to analytically model, which we leave as an open problem.
The pdf $P(x\vert T)$ calculated in Eq.~[\ref{Pxt}] is the conditional probability 
that an individual has achieved a career position $x$ by his/her given termination time $T$. 
Hence,  to obtain an ensemble pdf of career longevity $P(x)$ we must average over the pdf $r(T)$ of random termination
times $T$, 
\begin{equation}
P(x) = \int_{0}^{\infty} P(x \vert T) r(T) dT \ .
\label{EnsAve}
\end{equation}
We next make a suitable choice for $r(T)$. 
To this end, we introduce the hazard rate, $H(T)$, 
which  is the Bayesian probability that failure will occur at time $T + \delta T$, given that it has not yet occurred at
time $T$. This is written as 
$H(T) = r(T)/S(T) = - \frac{ 
\partial }{
\partial T} \ln S(T) $ , where $S(T) \equiv 1- \int_0^T r(t) \ dt $ is the probability of a career surviving until time
$T$. 
The exponential pdf of termination times, 
\begin{equation}
r(T) ={ x_{c}}^{-1} \text{exp}[-(T/x_{c})] \ ,
\label{expdist}
\end{equation}
has a constant hazard rate $ H(T) = \frac{1}{x_{c}}$, and thus assumes that external hazards are equally distributed over time. Substituting Eq.~[\ref{expdist}] into Eq.~[\ref{EnsAve}] and computing the integral, we obtain
 \begin{eqnarray}
P(x) &=& \frac{g(x)^{x-1}}{x_{c} (\frac{1}{x_{c}}+g(x))^{x}} 
\approx \frac{1}{g(x) x_{c}} \ e^{-\frac{x}{g(x)x_{c}}} \ .
\label{pxfin}
\end{eqnarray}
 Depending on the functional form of $g(x)$, the theoretical prediction given by Eq.~[\ref{pxfin}] is much different than the null model in which there is
no Matthew effect, corresponding to a constant progress rate $g(x) \equiv \lambda$ for each individual. 

Using the functional form given by Eq.~[\ref{gx}], we obtain a 
truncated power-law for the case of concave $\alpha <1$, resulting in a $P(x)$ that can  be
approximated by two regimes,
\begin{equation}
\label{longsol} P(x) \propto \left\{
\begin{array}{cl}
        x^{-\alpha} \ \ \ & x  \lesssim x_{c} \\
      e^{-(x/x_{c})}\  \ \ &  x \gtrsim x_{c} \ . \\
           \end{array}\right.
\end{equation}
Hence, our model predicts  a remarkable statistical regularity
which bridges the gap between very short and very long careers as a  result of  the concavity of $g(x)$ in early career development.

In the case of constant progress rate $g(x) \equiv \lambda$, the pdf $P(x)$ is exponential with a
characteristic career longevity $l_{c} = \lambda x_{c}$. 
 In the SI text we further consider the null model where the constant progress rate $\lambda_{i}$
of individual $i$ is
distributed over a given range. We  find again that $P(x)$ is exponential, which is quite different  than the prediction
given by Eq.~[\ref{longsol}].  Furthermore, we also develop a second model where the progress rate depends on a generic career trajectory $g(t)$ which peaks at a given year corresponding to the height of an individual's talent or creativity. 
We solve the time-dependent model in the SI text for a simple form of $g(t)$ which results in a $P(x)$ that is peaked around the maximum career length, in contrast to our empirical findings.

In order to account for aging effects,  another variation of this model could include a time-dependent $H(T)$. To
incorporate a non-constant $H(T)$ one can use a more general Weibull distribution for the pdf of termination times
\begin{equation}
r(T) \equiv  \frac{\gamma}{x_{c}}(\frac{T}{x_{c}})^{\gamma-1}\text{exp}[-(\frac{T}{x_{c}})^{\gamma}] \ ,
\end{equation}
 where $\gamma=1$ corresponds to the exponential case \cite{LifetimeStats}. In general, the hazard rate of the Weibull distribution is 
$H(T) \propto T^{\gamma-1}$, where $\gamma >1$ corresponds to an increasing hazard rate, and $\gamma <1$ corresponds to
a decreasing hazard rate.  We note that the timescale $x_{c}$ appears both in the definition of $g(x)$ in Eq.~[\ref{gx}] as a crossover between early and advanced career progress rates, and also 
as the timescale over which the probability of survival $S(T)$ approaches 0 in the case of $\gamma \geq 1$ in Eq.~[\ref{expdist}]. It is the appearance of the quantity $x_{c}$
in the definition of $S(T)$ that results in a finite exponential cutoff to the longevity distributions. Although the  timescales defined in $g(x)$ and $S(T)$ could be  different, 
 we observe only one timescale in the emprical data. Hence we assume here for simplicity that the two time scales are approximately equal.


\begin{figure}
\centerline{\includegraphics*[width=0.45\textwidth]{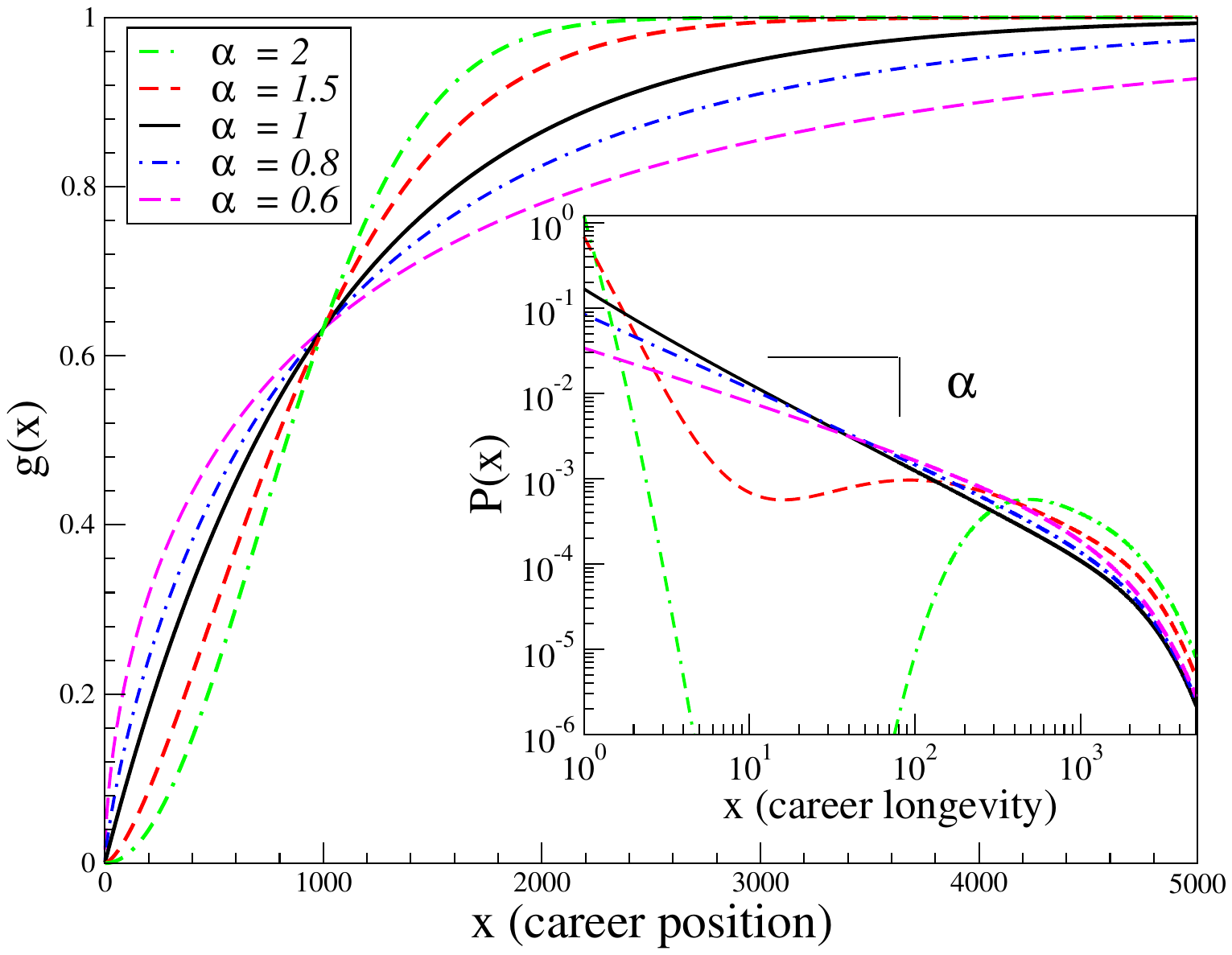}}
\caption{  Demonstration of the fundamental relationship between the progress rate $g(x)$ and the career longevity pdf
$P(x)$. 
 The progress rate $g(x)$ represents the probability of moving forward in the career to position $x+1$ from position
$x$. The small
 value of $g(x)$ for small $x$ captures the difficulty in making progress at the beginning of a career. 
 The progress rate increases with career position $x$, capturing the role of  the Matthew effect. We plot five $g(x)$
curves with fixed $x_{c} = 10^{3}$ and different values of the parameter $\alpha$. 
The  parameter $\alpha$ emerges from the small-$x$ behavior in $g(x)$ as the power-law exponent characterizing $P(x)$. 
{\it (Inset)} Probability density functions $P(x)$ resulting from inserting $g(x)$ with varying $\alpha$ into
Eq.~[\ref{pxfin}]. The value $\alpha_{c}\equiv 1$ separates two distinct types of longevity distributions. 
The distributions resulting from concave career development $\alpha <1$ exhibit monotonic statistical regularity
 over the entire range, with an analytic form approximated by the Gamma distribution $Gamma(x; \alpha, x_{c})$. The
distributions resulting from convex career development $\alpha >1$ exhibit bimodal behavior. In the bimodal case, one class of careers is 
stunted by the difficulty in making progress at the beginning of the career, analogous to a ``potential" barrier. The
second class of careers 
 forges beyond the barrier and is approximately centered around the crossover  $x_{c}$ on a log-log scale. \label{fig1} } 
\end{figure} 

From the theoretical curves plotted in the inset of Fig.~2, one observes that $\alpha_{c} =1$ is a special crossover value
for $P(x)$, between a bimodal $P(x)$ for $\alpha >1$, and a monotonically decreasing $P(x)$ for $\alpha <1$. 
 This crossover is due to the small $x$ behavior of the progress rate $g(x) \approx x^{\alpha}$ for $x < x_{c}$, which
serves as a ``potential barrier" that a young career must overcome. The width  $x_{w}$ of the potential barrier, defined
such that $g(x_{w})= 1/x_{c}$, scales as $x_{w}/x_{c} \approx x_{c}^{-1/\alpha}$. Hence, the value $\alpha_{c} =1$
separates convex progress $(\alpha >1)$ from concave progress $(\alpha <1)$ in  early career development. 

In the case
$\alpha >1$, 
one class of careers is stunted by the barrier, while the other class of careers excels, resulting in a  bimodal $P(x)$.
In the case $\alpha <1$, it is relatively more easy to make progress in the beginning of the career. It has been shown in Ref. \cite{ReedH} that random stopping times can explain power law pdfs in many stochastic systems that
arise in the natural and social sciences, with predicted exponent values $\alpha \geq 1$.  Our model provides a
mechanism which predicts truncated power-law pdfs with scaling exponents
 $\alpha \leq 1$, where the truncation is a requirement of normalization. 
Moreover, our model provides a quantitative meaning for the power-law exponent $\alpha$ characterizing the probability density function. 

\begin{figure}
\centerline{\includegraphics*[width=0.45\textwidth]{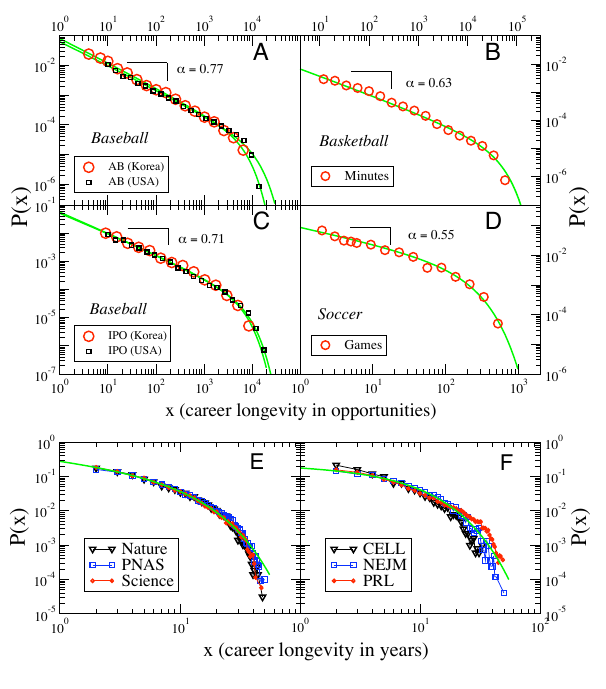}}
  \caption{ Extremely right-skewed pdfs $P(x)$ of career longevity in several high-impact scientific journals and 
several major sports leagues. We analyze data from American baseball (Major League Baseball) over the 84-year period
1920-2004, Korean Baseball (Korean Professional Baseball League) over the 25-year period 1982-2007, American basketball
(National Basketball Association and American Basketball Association) over the 56-year period 1946-2004, and English
soccer (Premier League) over the 15-year period 1992-2007, and several scientific journals over the 42-year period
1958-2000.  Solid curves represent least-squares best-fit functions corresponding to the functional form in Eq. [\ref{pxfin}]. {\it
(A) } Baseball fielder longevity
measured in at-bats (pitchers excluded): we find $ \alpha \approx 0.77$ , $x_{c}\approx 2500$ (Korea) and $x_{c} \approx
5000$ (USA). 
  {\it (B)} Basketball longevity measured in minutes played: we find $ \alpha \approx 0.63$, $x_{c} \approx 21000$
minutes.
  {\it (C)} Baseball pitcher longevity measured in innings-pitched measured in outs (IPO): we find $ \alpha
\approx 0.71$ , $x_{c} \approx 2800$ (Korea), and $ x_{c} \approx 3400$ (USA).
  {\it (D)} Soccer longevity measured in games played:  we find $ \alpha \approx 0.55$ , $x_{c} \approx$ 140 games.
{\it (E} and {\it F)} 
  High-impact  journals exhibit similar longevity distributions for the ``journal career length'' which we define as the
duration between an author's first and last paper in a particular journal. Deviations 
  occur for  long careers due to data set limitations (for comparison, least-square fits are
plotted in panel {\it (E)} with parameters $ \alpha \approx 0.40$, $x_{c} = 9$ years and in panel {\it (F)} with parameters $
\alpha \approx 0.10$, $x_{c} = 11$ years). These statistics are summarized in Table S2 of the SI. 
\label{fig2} }
\end{figure}

\section{Empirical Evidence} The two essential ingredients of our stochastic model, namely random forward
progress and random termination times corresponding to a stochastic hazard rate, 
are general and should apply in principle to many competitive professions. 
The individuals, some who are championed as legends and stars, are judged by their performances, usually on the basis of
measurable metrics for longevity, success, and prowess, which vary between professions. 

In scientific arenas, and in general, the metric for career position is difficult to define, even though there are many
conceivable metrics for career longevity and success \cite{ Hindex, physranking, music}. 
We compare author longevity within individual journals, which mimic an arena for competition, each with established
review standards that are related to the journal prestige. 
As a first approximation, the career longevity of a given author within a particular high-impact journal can be roughly measured as the duration
between an author's first and last paper in that journal, reflecting his/her ability to produce at the top tiers of
science. 
This metric for longevity should {\it not} be confused with the career length of the scientist, which is probably longer
than the career longevity within any particular journal. Following standard lifetime data analysis methods 
\cite{Huber98}, we collect ``completed" careers from our data set. The publication data we collect for each journal begins at year $Y_{0} = 1958$ for all journals except for CELL (for which $Y_{0} = 1974$),  
and ends at year $Y_{f} = 2008$. 

For each scientific career $i$, we calculate $\langle \Delta \tau_{i} \rangle$, the average
time between publications in a particular journal. A journal career which begins with a publication in year $y_{i,0}$
and ends with a publication in year $y_{i,f}$ 
is considered ``complete",  if the following two criteria are met: 
 $(a)  \ \ y_{i,f} \leq Y_{f} - \langle \Delta \tau_{i} \rangle$  and
 $(b)  \ \ y_{i,0} \geq Y_{0} +\langle \Delta \tau_{i} \rangle$.
 These criteria help eliminate from our analysis incomplete careers  which possibly began before $Y_{0}$ or ended after
$Y_{f}$.
 We then estimate the career length within journal $j$ as $L_{i,j}=y_{i,f} -y_{i,0} +1$, with a year allotted for
publication time, and do not consider careers with $y_{i,f} = y_{i,0}$.
 This reduces the size of each journal data set by approximately $25\%$ (see Table S1 and the SI text for a
description of data and methods). 

In Ref. \cite{CiteProd} we further analyze the scientific careers of the authors in these six journal data bases. 
In order to account for time-dependent and discipline dependent factors that affect both success and productivity measures, we 
develop normalized metrics for career success (``citation shares") and productivity (``papers shares").  
We also find further evidence of the Matthew effect by analyzing the  inter-publication time $\tau(x)$  which decreases with
increasing publication $x$ for individual authors within a given journal. 
Thus, we conclude that publication in a particular journal is facilitated by previous publications in the journal, corresponding to an increasing reputation within the given journal.
 Several other metrics  for quantifying career
success \cite{physranking, GenPubAuth},
such as the h-index \cite{Hindex} and generalizations \cite{Hgen, hindexResearchers, gappaper}, along with 
methods for removing time and discipline-dependent citation factors \cite{UnivCite} have been
analyzed in the spirit of developing unbiased rating systems for scientific achievement. 

In athletic arenas, the metrics for career position, success and success rate are more easy to define \cite{BB}. In
general, a career position in sports can be measured by the cumulative number of in-game opportunities a player has
obtained. In baseball, we define an opportunity as an {\it ``at bat"} (AB) for batters, and an  {\it ``inning pitched in
outs"} (IPO) for pitchers, while in basketball and soccer, we define the metrics for opportunity as {\it ``minutes
played"} and {\it ``games played "}, respectively. Methods from network science  have recently been used to analyze measures for career success
in professional Tennis \cite{Tennis}.

In Fig.~3 we plot the distributions of career longevity for  $20,000$ professional athletes in four distinct leagues and roughly
$400,000$ scientific careers in six distinct journals  (data is publicly available at \cite{SportsStats,journallimiations}). 
We observe universal statistical regularity corresponding to $\alpha <1$ in the career longevity distributions for three
distinct sports and several high-impact journals (see Table S2 for a summary of  least squares parameters). 
The disparity in career lengths indicates that it is very difficult to sustain a competitive professional career, with
most individuals making their debut and finale over a relatively short time interval. 
For instance,  we find that roughly 3\% of baseball pitchers have a career length in the MLB of one inning pitched or less, while we also find that
roughly 3\% of basketball players have a career length in the NBA of less than 12 in-game minutes.
Yet, despite the relatively high frequency of short careers, there are also instances of careers that are extremely long, corresponding to roughly the entire productive lifetime of the individual. The statistical regularity which bridges the gap between the ``one-hit wonders'' and the ``iron horses'' indicates that there are careers of every length between the minimum and the maximum career length, with a smooth and monotonic relation quantifying the relative frequencies of the careers in between. 
Furthermore, we find that stellar careers are not anomalies, but rather, as predicted by our model, the outcome of the cumulative advantage in competitive professions. The properties of the cumulative advantage process are also compounded by an individual's  ``sacred spark'' factor \cite{AccumAdvProdDiff} which accounts for his/her relative level of talent and/or professional drive, which also factors into career longevity. 

The exponential cutoff in $P(x)$
that follows after the crossover value $x_{c}$, arises from the finite human lifetime, and is reminiscent of any real
system where there are finite-size effects that dominate the asymptotic behavior. The scaling regime is less pronounced in the curves for journal longevity. This results from the granularity of our data set, which records publications by
year only. A
 finer time resolution (e.g. months between first and last publication) would likely reveal a larger scaling regime. However,
regardless of the scale, one observes the salient feature of there being a large disparity between the frequency of long
and short careers.

In science, an author's success metric can be quantified by the
total number of papers or citations in a particular journal. 
Publication careers have the important property that the impact of scientific work is time dependent. Where many papers 
become outdated as the scientific body of knowledge grows, there are instances where ``late-blooming" papers make significant impact
a considerable time after publication \cite{110PhsRev}.
Accounting for the time-dependent properties of citation coutns, in  \cite{CiteProd} we  find that the pdf of total number of normalized
citation shares for a particular author in a single journal over his/her entire career follows the asymptotic power law $P(z)dz\sim 
z^{-2.5}dz$ for the six journals analyzed here. 

 In sports, however, career accomplishments do not wax or wane with time. In Fig.~4 we plot the
pdf $P(z)$ of career success $z$ for common metrics in baseball and basketball.
Remarkably, the power-law regime for $P(z)$ is governed by a scaling exponent which is approximately equal to the scaling exponent of the longevity pdf $P(x)$.
In the SI, we show  that the pdf $P(z)$ of career success $z$ follows
directly from a simple Mellin convolution of the pdf $P(x)$ for longevity $x$ and the pdf $P(y)$ of prowess $y$.

\begin{figure}
\centerline{\includegraphics*[width=0.45 \textwidth]{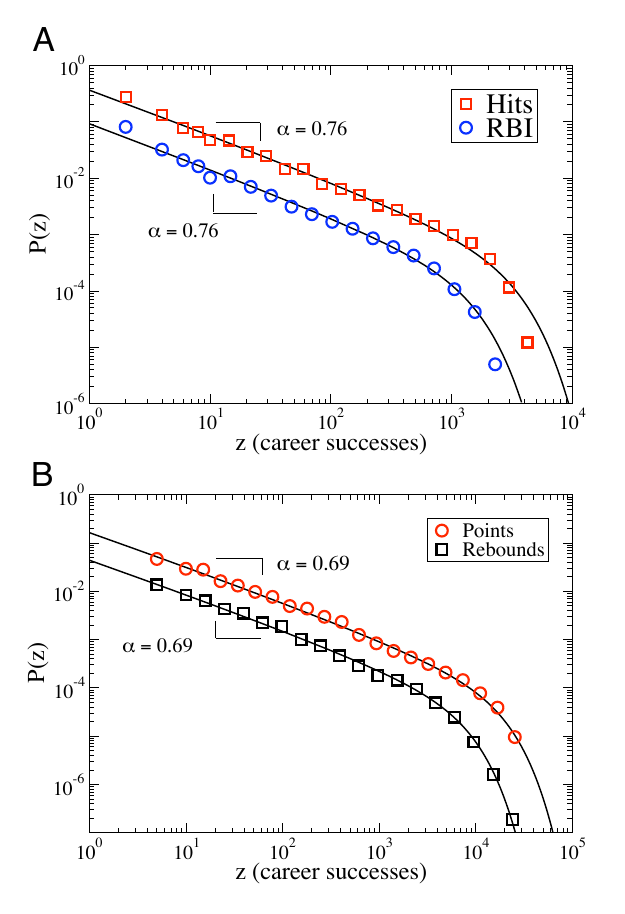}}
  \caption{  Probability density function $P(z)$ of common metrics for career success, $z$.
   Solid curves represent best-fit functions corresponding to Eq. [\ref{pxfin}].
  {\it(A)} Career batting statistics in American baseball: $x_{c}^{Hits}\approx 1200$, $x_{c}^{RBI}\approx 600$, (RBI =
Runs Batted In).  
  {\it (B)} Career statistics in American basketball: $x_{c}^{Points}\approx 8000$, $x_{c}^{Rebounds}\approx 3500$.
  For clarity, the top set of data in each plot has been multiplied by a constant factor of four in order to separate
overlapping data.}
 \label{fig3}
\end{figure}

The Gamma pdf
 $ P(x) \equiv Gamma(x; \alpha, x_{c})\propto x^{-\alpha}\ e^{-x/x_{c}}$ is commonly employed in statistical modeling,
and can be
 used as an approximate form of  Eq.~[\ref{longsol}].
One advantage to the gamma pdf is that it can be inverted in order to study extreme statistics corresponding to rare
stellar careers. 
In the SI and in \cite{BB3}, we further analyze the relationship between the  extreme statistics of the Gamma pdf and the
selection processes for {\it Hall of Fame} museums. 
In general, the statistical regularity of these distributions allows one to establish robust milestones, which could be used for setting
the corresponding financial rewards and pay scales, within a particular profession.  Interestingly, we also find in \cite{BB3} that the pdfs for career success in MLB are stationary even 
if we quantitatively remove the time-dependent factors that can relatively inflate or deflate measures for success. This stationarity implies that the right-skewed statistical regularity we observe  in $P(z)$ arises from both the intrinsic talent and the longevity of professional athletes, and does not result from 
 changes in technology, economic factors, training improvements, etc. In the case of MLB, this detrending method allows one to compare the 
 accomplishments of baseball players across historical eras, and in particular, can help to interpret and quantify the relative achievements of 
 players from the recent ``steroids era."

In summary, a wealth of data recording various facets of social phenomena have become available in recent years,
allowing scientists to search for universal laws that emerge from human interactions \cite{CompSocialScience}.
Theoretical models of social dynamics, employing methods from statistical physics, have provided significant insight
into the various mechanisms that can lead to emergent phenomena \cite{SocialDynamics}. 
An important lesson from complex system theory
is that oftentimes the details of the underlying mechanism do not affect the macroscopic emergent phenomena. 
 For baseball players in Korea and the United States, we observe remarkable similarity between the pdfs of career
longevity (Fig.~3) and the pdfs of prowess  (Fig.~S1), despite these players belonging to completely distinct leagues. 
  This fact is consistent with the hypothesis that 
universal stochastic forces  govern career development in science, professional sports and presumably in a large class
of competitive professions.

In this paper we demonstrate strong empirical evidence for  universal statistical laws that describe career progress in competitive professions.
Universal phenomena also occur in many other social complex systems where regularities arise despite the complexity of the human interactions and the spatio-temporal dynamics \cite{socpowlaw,PAsex,MEJN,
econpowlaw,Barabasibursts,wattsSW,mobility,Emails,SornettePNAS,linking,FOMC,MetroHopper1, GravityHighway, HumanMobility2}.
Stemming from the simplicity of the assumptions, the stochastic model developed in this paper could concievably apply elsewhere in
society, such as the duration of
both platonic and romantic friendships. Indeed, long relationships are harder to break than short ones, with random
factors inevitably terminating them forever. Also, supporting evidence for the applicability of this model can be found in the
similar truncated power-law pdfs with $\alpha<1$, that describe the dynamics of connecting within online social
networks
\cite{linking}.

We thank P. Krapivsky, G. Viswanathan, G. Paul, F. Wang, and J. Tenenbaum for insight and helpful comments. 
AMP and HES  thank the ONR and DTRA for financial support, WSJ was supported by the Basic Science Research Program through the
National Research Foundation of Korea (NRF) funded by the Ministry of Education, Science and Technology grant 2010-0021987, and J.-S. Yang thanks grant NRF-2010-356-B00016 for financial support. 

\section{Appendix: the Spatial Poisson Distribution} The master equation for the evolution of $P(x,N)$ is
\begin{eqnarray}
\label{ME} 
P(x+1,N+1) &-& P(x+1,N) \nonumber \\ 
= f(x)P(x,N) &-& f(x+1)P(x+1,N) \ ,
\end{eqnarray}
with initial condition,
\begin{eqnarray}
P(x+1,0) = \delta_{x,0} \ .
\label{IC}
\end{eqnarray}
Here $f(x)$ represents the probability that an employee obtains another future opportunity given his/her resume at
career position $x$. 
We next write the discrete-time discrete-space master equation in the continuous-time discrete-space form, 
\begin{eqnarray}
   \frac{
\partial P(x+1,t)}{
\partial t } =
    g(x)P(x,t) &-& g(x+1)P(x+1,t ) \ ,
\end{eqnarray}
where $g(x) = f(x)/\delta t$ and $t= N \delta t$ \ (for an extensive discussion of master equation formalism 
see Ref.~\cite{firstpassage}). Taking  the Laplace transform of both sides one obtains, 
  \begin{eqnarray}
 \lefteqn{ sP(x+1,s)-P(x+1,t=0)=} \nonumber \\
&& \ \ \ \ g(x)P(x,s)-g(x+1)P(x+1,s) \ .
\end{eqnarray}
From the initial condition in Eq.~[\ref{IC}] we see that the second term above vanishes for $x\geq 1$. Solving for
$P(x+1,s)$ we obtain the recurrence  equation
\begin{eqnarray}
 P(x+1,s) = \frac{g(x)}{s+g(x+1)} P(x,s) \ .
\end{eqnarray}
If the first derivative $\frac{d}{dx}g(x)$ is relatively small, we can replace $g(x+1)$ with $g(x)$ in the equation above.
Then, one can verify the ansatz
\begin{eqnarray}
\label{Pxslapl}
 P(x,s) = \frac{g(x)^{x-1}}{(s+g(x))^{x}} \ ,
\end{eqnarray} 
which is the Laplace transform of the spatial Poisson distribution  $P(x,t ; \lambda = g(x))$ as in \cite{spatialPoisson}. 
The Laplace transform is  defined as $L\{f(t)\} = f(s) = \int_{0}^{\infty} dt f(t) e^{-s t} $. Inverting the
transform we obtain
\begin{eqnarray}
 P(x,t) = \frac{e^{-g(x)t}(g(x)t)^{x-1}}{(x-1)!} \ .
 \label{sPxt}
\end{eqnarray} 
Hence, Eq.~[\ref{sPxt}] corresponds to the pdf of final career position $x$ observed at a particular time $t$. Since not all careers last the same length of time,
we define the time $t \equiv T$ to be a conditional stopping time  which characterizes a  given subset of careers that lasted a time duration $T$. We
average over a distribution $r(T)$ of 
stopping times to obtain the empirical longevity pdf $P(x)$ in Eq.~[\ref{pxfin}], which is equivalent to  Eq.~[\ref{Pxslapl}], so that $P(x)$ is comprised of careers with varying $T$.

 \renewcommand{\theequation}{S\arabic{equation}
}
  \renewcommand{\thefigure}{S\arabic{figure}}
    \renewcommand{\thetable}{S\arabic{table}}

      \setcounter{equation}{0}  
  \setcounter{figure}{0}
    \setcounter{table}{0}
    \setcounter{section}{0}
    \begin{center}
  \section*{Supporting Information}
  \end{center}

 \section{Data and methods} 
The publication data analyzed in this paper was downloaded  from {\it ISI Web of Knowledge} in May 2009. We restrict our
analysis to publications termed as ``Articles", which excludes reviews, letters to editor, corrections, etc. 
Each article summary includes a field for the author identification consisting of a last name and first and middle
initial (eg. the author name John M. Doe would be stored as ``Doe, J" or ``Doe, JM" depending on the author's
designation). From these fields, we collect the career works of individual authors within a particular journal together,
and analyze metrics for career longevity and success.\\

For author $i$ we combine all articles in journal $j$ for which he/she was listed as coauthor. The total number of
papers for author $i$ in  journal $j$ over the 50-year period is $n_{i}$. 
Following  methods from lifetime statistics \cite{Huber98s}, we use a standard method to isolate ``completed"
careers from our data set which begins at year $Y_{0}$ 
and ends at year $Y_{f}$. For each author $i$, we calculate $\langle \Delta \tau_{i} \rangle$, the average time $\Delta
\tau_{i}$ between successive publications in a particular journal. A career which begins with the first recorded
publication in year $y_{i,0}$ and ends with the final recorded publication in year $y_{i,f}$ 
is considered ``complete",  if the following two criteria are met: 
\begin{itemize}
\item[(1)] $  y_{i,f} \leq Y_{f} - \langle \Delta \tau_{i} \rangle$  \\
\item[(2)] $ y_{i,0} \geq Y_{0} +\langle \Delta \tau_{i} \rangle$.\\
 \end{itemize}
This method estimates that the career begins in year $y_{i,0}-\langle \Delta \tau_{i} \rangle$ and ends in year  $y_{i,f}+\langle \Delta \tau_{i} \rangle$. 
 If either the estimated beginning or ending year do not lie within the range of the data base, than we discount the
career as incomplete to first approximation. Statistically, this 
 means that there is a significant probability that this author published before $Y_{0}$ or will publish after $Y_{f}$. 
 We then estimate the career length within journal $j$ as $L_{i,j}=y_{i,f} -y_{i,0} +1$, and do not consider careers
with $y_{i,f} = y_{i,0}$.
 This reduces the size of the data set by approximately $25\%$ (compare the raw data set sizes $N$ to the pruned 
 data set size $N^{*}$ in Table \ref{table:journals}).

There are several potential sources of systematic error in the use of this database:
\begin{itemize}
\item[(i)] Degenerate names $\rightarrow$ increases career totals. Radicchi {\it et al.} \cite{physranking2}  observe
that this method of concatenated author ID leads to a pdf $P(d)$ of degeneracy $d$ which scales as $P(d) \sim d^{-3}$.
\item [(ii)] Authors using middle initials in some but not all instances of publication $\rightarrow$ decreases career
totals.
\item [(iii)] A mid-career change of last name $\rightarrow$  decreases career totals.
\item [(iv)] Sampling bias due to finite time period. Recent young careers are biased toward short careers. 
Long careers located towards the beginning $Y_{0}$ or end $Y_{f}$ of the database are biased towards short careers.
\end{itemize}

 \section{A Robust Method for Classifying Careers   \label{sec:robust} } Professional sports leagues are geared around
annual  championships that celebrate the accomplishments of teams over a whole season. On a player level, professional
sports leagues annually induct retired players into ``halls of fame" in order to celebrate and honor stellar careers. 
Induction immediately secures an eternal legacy for those that are chosen. However, there is no standard method for
inducting players into a {\it Hall of Fame}, with subjective and political factors affecting the induction process. 
In \cite{BB32} we quantitatively normalize seasonal statistics so to remove time-dependent factors that
influence success. This provides a framework for comparing career statistics across historical eras.

In this section we propose a generic and robust method for measuring careers. 
We find that the pdf for career longevity can be approximated by the gamma distribution,
\begin{equation}
Gamma(x; \alpha, x_{c}) = \frac{x^{-\alpha}e^{-x/x_{c}}}{x_{c}^{1-\alpha}\Gamma(1-\alpha)} \ ,
\end{equation}
with moments $\langle x^{n}\rangle = x_{c}^{n} \frac{\Gamma(1-\alpha+n)}{\Gamma (1-\alpha)},$ where we restrict our
considerations to the case of 
$\alpha \leq 1$, with $x_{c} >>1$. This distribution allows us to calculate the extreme value $x^{*}$ such that only  $f$ percentage of players  exceed this value according to the pdf $P(x)$,
\begin{eqnarray}
f &=& \int_{x^{*}}^{\infty} \frac{x^{-\alpha}e^{-x/x_{c}}}{x_{c}^{1-\alpha}\Gamma(1-\alpha)}dx \noindent \\
&=& \frac{\Gamma[1-\alpha,\frac{x^{*}}{x_{c}}]}{\Gamma(1-\alpha)} = Q[1-\alpha, \frac{x^{*}}{x_{c}}] \ ,
\end{eqnarray}
where $\Gamma[1-\alpha,\frac{x^{*}}{x_{c}}]$ is the incomplete gamma function and $Q[1-\alpha, \frac{x^{*}}{x_{c}}]$ is
the regularized gamma function. This function can be easily inverted numerically using computer packages, e.g. {\it
Mathematica}, which results in  the statistical benchmark 
\begin{equation}
x^{*}= x_{c} \ Q^{-1}[1-\alpha,f].
\end{equation}
 In \cite{BB32} we use the maximum
likelihood estimator (MLE) for the Gamma pdf to estimate the parameters $\alpha$ and $x_{c}$ for each pdf. The values we
obtain using MLE are systematically smaller for $\alpha$ values and  for $x_{c}$ values, but the relative
differences are negligible. 

In Table \ref{table:stats} we provide statistical benchmarks $x^{*}$ corresponding to career longevity and career metrics
for several sports. For the calculation of each $x^{*}$  we use the parameter values $\alpha$  and $x_{c}$ calculated
from a least-squares fit to the empirical pdf $P(x)$ using the functional form of Eq.~[5], and the  
significance level value $f$ calculated from historical
induction frequencies in the American Baseball Hall of Fame (HOF) in Cooperstown, NY USA. The baseball HOF has inducted
276 players out of the 14,644 players that exist in Sean Lahman's baseball database between the years 1879-2002, which 
corresponds to a fraction $f\equiv0.019$. It is interesting to note that the last column, $\frac{x^{*}}{\sigma}
\equiv \beta\approx 3.9$ for all the gamma distributions analyzed. 
 This approximation is a consequence of the universal scaling form of the gamma function $Gamma(x) \equiv
U(x/x_{c})$, where the standard deviation $\sigma$ of the Gamma pdf has the simple relation $\sigma =
x_{c}\sqrt{1-\alpha}$.
Hence,  for a given $f$ and $\alpha$, the ratio 
\begin{equation}
x^{*} / \sigma = \frac{Q^{-1}[1-\alpha,f]} {\sqrt{1-\alpha}}
\end{equation} is independent of $x_{c}$.
Furthermore, this approximation is
 valid for all statistics
in MLB since $\alpha$ is approximately the same for all pdfs analyzed. Thus, the value $x^{*} \approx 4 \sigma$
is a robust approximation for determining if a player's career is stellar at the $f \approx 0.02$ significance level. 
The highly celebrated milestone
of 3,000 hits in baseball corresponds to the value  $ x^{*} = 1.26 \ \beta \sigma_{hits}$. Only 27 players have exceeded
this benchmark in their professional careers, while only 86 have exceeded the arbitrary 2,500 benchmark. 
Hence, it makes sense to set the benchmark for all milestones at a value of $x^{*} = \beta \sigma$ corresponding to
each distribution of career metrics. \\

We  check for consistency by comparing the extreme threshold value $x^{*}$ calculated using the gamma distribution with
the value
$x_{d}^{*}$ derived from the database of career statistics. Referring to the actual set of all baseball players from
1871-2006, to achieve a fame value $f_{d} \approx 0.019$ with respect to hits, one should set the statistical benchmark
at $x_{d}^{*} \approx 2250$, which account for 146 players 
(this assumes that approximately half of all baseball players are not pitchers, who we exclude from this calculation of
$f_{d}$). The value of $x_{d}^{*}\approx 2250$ agrees well with the value calculated from the gamma distribution, $x^{*}
\approx 2366$. Of these 146 players with career hit tallies greater than 2250, there are 126 players who have been eligible for at
least one induction round, and 82 of these players have been successfully inducted into the American baseball hall of
fame. Thus, a player with a career hit tally above $x^{*} \approx x_{d}^{*}$ has a $65\%$ chance of being accepted,
based on just those merits alone. Repeating the same procedure for career strikeouts obtained by pitchers in baseball we
obtain the milestone value $x_{d}^{*}\approx 1525$ strikeouts, and for career points in basketball we obtain the value
$x_{d}^{*}\approx 16,300$ points. Nevertheless, the overall career must be taken into account, which raises the bar, and
accounts for the less than perfect success rate of being voted into a hall of fame, 
given that a player has had a statistically stellar career in one statistical category.

\section{Career Metrics} 

In Fig.~4 we plot common career metrics for success in American baseball and American basketball.  
Note that the exponent $\alpha$ for the pdf $P(z)$ of total career successes $z$ is approximately equal to the exponent
$\alpha$ for the pdf $P(x)$ of career longevity $x$ (see Table~\ref{table:stats}).  
In this section, we provide a simple explanation for the similarity  between the  power law exponent for career
longevity (Fig.~2) and the  power law exponent  for career success (Fig.~4). 

Consider a distribution of longevity that is power law distributed, $P(x) \sim x^{-\alpha}$ for the entire range $1 \leq
x \leq x_{c} < \infty $.
 The cutoff $x_{c}$ represents the finiteness of human longevity, accounted for by the exponential decay in Eq.~[7].
Also, assume that the prowess $y$ has a pdf $P(y)$ which is characterized by a mean and standard deviation, which
represent the talent level among professionals (see Ref. \cite{BB} for the corresponding prowess distributions in major
league baseball). 
In the first possible case, the distribution is right-skewed and approximately exponential (as in the case of
home-runs). 
In other cases, the distributions are essentially Gaussian. Regardless of the distribution type, the prowess pdfs $P(y)$
are confined to the domain $ \delta \leq y \leq  1 $, where $\delta > 0$.

Assume that in any given appearance, a person can apply his/her natural prowess towards achieving a success, independent
of past success. Although prowess is refined over time,  this  should not substantially alter our 
demonstration. Since  not all professionals have the same career length,  the career totals are in fact a combination of
these two distributions as in their product. Then the career success total  $z=xy$ has the distribution,
\begin{eqnarray}
P(z = xy) &=& \int \int dy \ dx \ P(y)P(x) \delta (xy-z) \nonumber \\
&=&\int \int dy \ dx P(y)P(x) \delta (x(y-z/x)) \nonumber \\
&=& \ \int dx \ P(\frac{z}{x} ) P(x) \frac{1}{x}  \ .
\label{Fz}
\end{eqnarray}
This integral has three domains (Ref.~\cite{2randv}),
\begin{eqnarray}
P(z) \propto \left\{
\begin{array}{cl}
        \int_{1}^{z/\delta} dx \ P(\frac{z}{x}) x^{-(\alpha+1)} \ , \ & \delta < z < 1\\
\nonumber
        
        \int_{z}^{z/\delta} dx \ P(\frac{z}{x})  x^{-(\alpha+1)}  \ , \ & 1 < z < x_{c} \delta\\
        
        \int_{z}^{x_{c}} dx \ P(\frac{z}{x})  x^{-(\alpha+1)}  \ , \ & x_{c} \delta < z < x_{c} \ . \\
           \end{array}\right.
\end{eqnarray}
The first regime $ \delta < z < 1$ is irrelevant, and  is not observed since z is discrete in the cases analyzed here.
For the first case of an exponentially distributed prowess, 
\begin{eqnarray}
P(z) \propto \left\{
\begin{array}{cl}
        z^{-\alpha} \ , \ & 1 < z <x_{c} \ \  \  \delta \\ 
        z^{-\alpha} \ \text{exp}(-z/\lambda x_{c})\ , \  \  \ & x_{c} \delta< z < x_{c} \ .
           \end{array}\right.
           \label{Fz1}
\end{eqnarray}

In Ref.~\cite{BBs} we mainly observe the exponential tail in the home-run distribution, as the above form suggests in the
regime $x_{c} \delta< z < x_{c}$, resulting from $\delta \approx 0$ for the right-skewed home-run prowess distribution.
However, in the case for a normally distributed prowess, the power law behavior of the longevity distribution is
maintained for large values into the career success distribution $P(z)$,  as $x_{c} \delta > 10^{3}$.
\begin{eqnarray}
P(z) \propto \left\{
\begin{array}{cl}
        z^{-\alpha} \ , \ & 1 < z < x_{c} \delta\\
        z^{-\alpha} e^{-(\frac{z}{\sigma x_{c}})^{2}/2}\ , \ & x_{c} \delta< z < x_{c} \ .
           \end{array}\right.
           \label{Fz2}
\end{eqnarray}
Thus, the main result of this demonstration is that the distribution $P(z)$ maintains the power law exponent $\alpha$ of
the career-longevity distribution, $P(x)$, when the prowess is distributed with a characteristic mean and standard
deviation. 
This result is also  demonstrated with the simplification of representing the prowess distribution $P(y)$ as an
essentially uniform distribution over a reasonable domain of $y$, which simplifies the integral in Eq.~(\ref{Fz}) while
maintaining the inherent power law structure.

\begin{figure}
\centerline{\includegraphics*[width=0.45\textwidth]{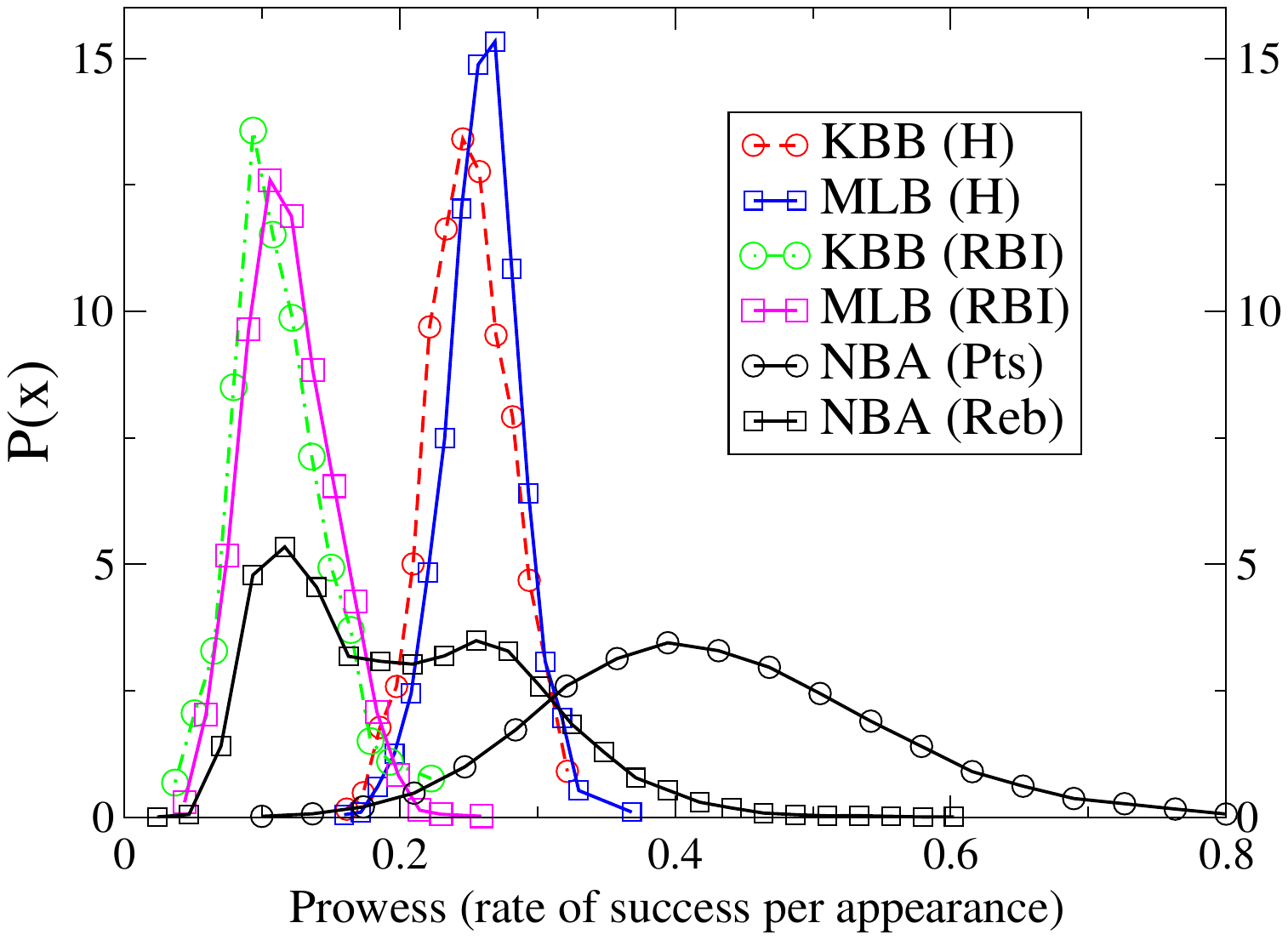}}
  \caption{Probability density functions of seasonal prowess for several career metrics. 
  Each pdf is  normally distributed, except for the bimodal curve for rebound prowess, NBA (Reb.). 
The bimodal distribution for  Rebound prowess reflects the specialization in player positions in the sport of
basketball. 
Furthermore, note the remarkable similarity in the distributions between American (MLB) and Korean (KBB) baseball
players. }\label{fig4} 
\end{figure}

 In Fig.~\ref{fig4} we plot the prowess distributions that correspond to the career success distributions plotted in
Fig.~4. It is interesting that the competition level based on the distributions of prowess indicates that Korean and
American baseball are nearly equivalent. Also, note that the prowess distributions for rebounds per minute are bimodal,
as the positions of players in basketball are more specialized. 

 \section{A null model without the Matthew effect}
 
 In this section, we compare the predictions of our theoretical model with the predictions of
a theoretical model which does not incorporate the Matthew effect.
Since the Matthew effect implies that the progress rate $g(x)$
increase with career position $x$, we analyze the more simple model where for each individual $i$ the progress rate
$g_{i}(x)$
is constant, 
\begin{equation}
g_{i}(x) \equiv \lambda_{i} \ .
\end{equation}
The solution to the conditional longevity pdf $P(x \vert \lambda_{i})$ is still given  by Eq.~[5], taking the form
\begin{equation}
P(x \vert \lambda_{i}) = \frac{\lambda_{i}^{x-1}}{x_{c} (\frac{1}{x_{c}}+\lambda_{i})^{x}}  \approx \frac{1}{\lambda_{i}
x_{c}} \ e^{-\frac{x}{\lambda_{i} x_{c}}} \ ,
\end{equation}
which is an exponential pdf, with a characteristic career length $l_{c} \equiv \lambda_{i} x_{c}$. 
 Hence, this null model corresponds to a career progress mechanism wherein
intrinsic ability, which is incorporated into the relative value of $\lambda_{i}$, is the dominant factor.
In order to calculate the longevity pdf $P(x)$ which incorporates a distribution of intrinsic abilities across the
population of individuals, we average over the conditional pdfs using a pdf $P(\lambda)$ that we assume is well-defined
by a mean $\overline \lambda$ and standard deviation $\sigma$, consistent with what we observe for the seasonal
prowess pdfs shown in Fig.~S1. In the case of $P(\lambda)= Normal(\overline \lambda, \sigma)$, then 
\begin{equation}
P(x) = \int_{0}^{1}P(\lambda) P(x \vert \lambda)  d\lambda  \  \equiv \int_{0}^{1}
\frac{e^{-(\lambda-\overline{\lambda})^{2}/2\sigma^{2}}}{\sqrt{2\pi
\sigma^{2}}} 
P(x \vert \lambda) d\lambda \ .
\end{equation}
For the sake of providing an analytic result, we replace $P(\lambda)$ by a uniform distribution,
 \begin{equation}
\label{uniformP} P(\lambda) \approx \left\{
\begin{array}{cl}
       0 \ , & \vert \lambda - \overline\lambda   \vert  > 2 \sigma  \\
      \frac{1}{4\sigma} \ , & \vert \lambda - \overline\lambda  \vert  \leq 2 \sigma \  , \\ 
           \end{array}\right.
\end{equation}
which does  not change the overall result. The integral in Eq. (S10) then becomes,
\begin{eqnarray}
P(x) &\approx& \frac{1}{4\sigma}  \int_{\lambda- 2 \sigma}^{\lambda+ 2 \sigma} \frac{ d\lambda }{\lambda
x_{c}} \ e^{-\frac{x}{\lambda x_{c}}} \noindent \\
 &=&  \frac{1}{4\sigma x_{c}} [ \Gamma(0,\frac{x/x_{c}}{\overline\lambda + 2
\sigma}) - \Gamma(0,\frac{x/x_{c}}{\overline\lambda - 2 \sigma})]  \noindent \\
&\approx& e^{-x/\overline\lambda x_{c}} \ ,
 \end{eqnarray}
 for $1 > \overline \lambda > 2 \sigma$, where the last approximation corresponds to a relatively small $\sigma$. Thus,
we find that even with a reasonable dispersion in the constant progress rates $\lambda$ in  a population of individuals,
the pdf $P(x)$ is still exponential. Hence, our theoretical model cannot explain the empirical non-exponential  form of
$P(x)$  unless we incorporate the Matthew effect using $g(x)$ that
increase with $x$.

 \section{A null model with time-dependent career trajectory}
 
\begin{figure}[t]
\centerline{\includegraphics*[width=0.45\textwidth]{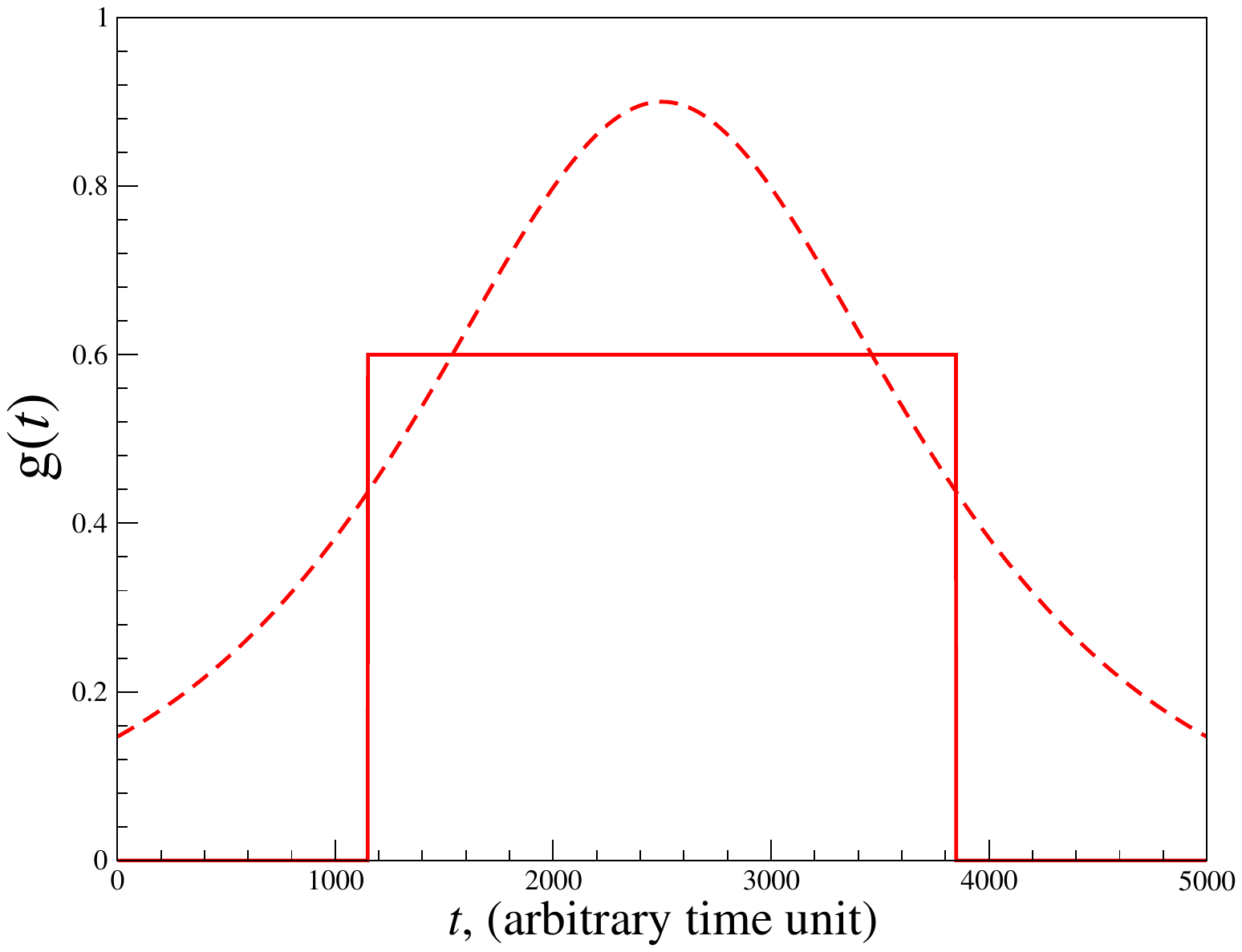}}
  \caption{A graphical  illustration of a hypothetical career progress trajectory $g(t) = a \ \text{sech} [(t-t^{*})/w]$ (dashed red line), with amplitude $a=0.9$, 
  peak time $t^{*}= 2500$, and width $w=1000$,  in arbitrary time units. As an approximation, in order to provide an analytic solution to the model, we approximate $g(t)$ by a uniform plateau function $g(t) \approx \gamma [H(t-t_{1})-H(t-t_{2})]$ (solid red line), as in Eq. (\ref{uniformgt}), where $H(t)$ is the standard Heavyside step function. }\label{figS2} 
\end{figure}

In this section, we develop a career progress model where the progress rate $g(t)$ is time-dependent instead of being position-dependent $g(x)$, as in the previous sections.  We use a time dependent career trajectory to capture the non-monotonic peaks in key productivity factors, e.g. creativity and talent, that are observed for various creative careers \cite{CareerTrajectorys}.    In Fig. \ref{figS2} we show a generic $g(t)$ which peaks at a variable time $t^{*}$, and has
 an amplitude $a$ related to the individual's underlying talent.  The regime in which $g(t)$ is increasing reflects the learning curve associated with a difficult endeavor, whereas the regime in which $g(t)$ is decreasing reflects e.g. aging factors and the upper limit to the finite resources which facilitate improvement. 

In analogy to Eq. [10], the master equation for the evolution of career progress is
\begin{equation}
\frac{\partial P(x+1,t)}{ \partial t } = g(t)P(x,t) - g(t)P(x+1,t ) \ ,
\label{ME2}    
\end{equation}
 where $g(t)$ is an arbitrary function which quantifies the forward progress rate at time $t$.
 To solve for $P(x,t)$, we define the ``integrated time'' $\tau$ given by,
\begin{equation}
\tau \equiv \int_{0}^{t} dt' g(t') \ .
\end{equation}
Hence, we write Eq. (\ref{ME2}) as,
\begin{equation}
\frac{\partial P(x+1,\tau)}{ \partial \tau } = P(x,\tau) - P(x+1, \tau) \ ,  
\end{equation}
which along with the initial condition $P(x+1, \tau) = P(x+1,t) = \delta_{x,0}$, has the solution,
\begin{equation}
P(x,\tau) = \frac{e^{-\tau} \tau^{x-1}}{(x-1)!} \ .
\end{equation}
As previously described in the main text, we obtain the unconditional probability density function $P(x)$ of career longevity $x$ from the conditional pdf  $P(x | T) = P(x,t \equiv T)$ using a pdf of random termination times $r(T)$, 
\begin{equation}
P(x) = \int_{0}^{\infty} P(x | T) r(T) dT \ ,
\label{Pave}
\end{equation}
where we use the exponential pdf $r(T) = x_{c}^{-1} \exp[-(T/x_{c})]$ for the demonstration of a career termination model with constant hazard rate, corresponding to the laplace transform of $P(x | T)$ in the variable $s = 1/x_{c}$. The integral in Eq. (\ref{Pave}) is typically difficult to calculate given the time-dependence of the  progress rate.\\

\begin{figure}[t]
\centerline{\includegraphics*[width=0.45\textwidth]{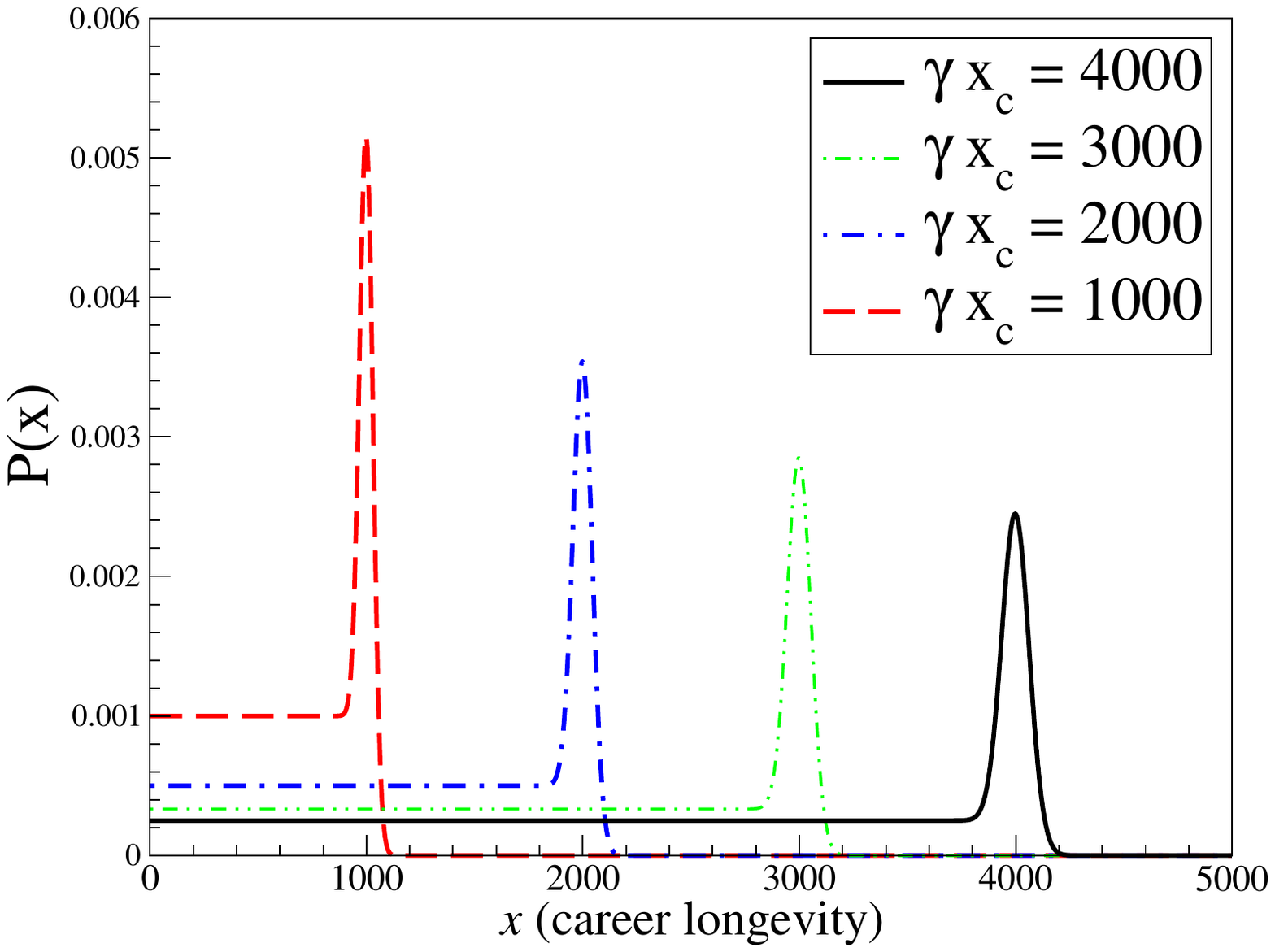}}
  \caption{ Exact solutions for $P(x)$ with time-dependent career trajectory $g(t)$ defined in Eq. (\ref{Pxexact}), for the case of $t_{1}=0$, $x_{c}=t_{2}$, and $\gamma x_{c} = \{1000, 2000, 3000, 4000\}$.  }\label{figS3} 
\end{figure}

 Simonton \cite{CareerTrajectorys} finds that the annual productivity of creative products or ideas has a trajectory that is peaked around a given characteristic time $t^{*}$ into a given profession. This peak is determined by  two model parameters quantifying ``ideation'' and ``elaboration'' rates, and two additional parameters quantifying initial creative potential and the age at career onset. 
To demonstrate the solution to our null model, we use an simplified functional form for $g(t)$ corresponding to a uniform distribution over the interval $t \in [t_{1},t_{2}]$,
 \begin{equation}
\label{uniformgt} g(t) \approx \left\{
\begin{array}{cl}
       0 \ , & t < t_{1}  \\
      \gamma \ , &  t \in [t_{1},t_{2}]  \\ 
       0  \ , & t > t_{2}  \ , 
           \end{array}\right.
\end{equation}
where $t_{1}$ is the ``breakout'' year of the career, $t_{2}$ corresponds to  the year in which the individual's productivity declines rapidly, and $0 \leq \gamma \leq 1$ is the intrinsic potential or talent of the given individual, and the time duration  $t_{2}-t_{1}$ is the precocity of the given individual. Hence, the corresponding integrated time $\tau$ is given by
 \begin{equation}
\label{Itgt} \tau \equiv   \int_{0}^{t} dt' g(t')  = \left\{
\begin{array}{cl}
       0 \ , & t < t_{1}  \\
       \gamma(t-t_{1}) \ , &  t \in [t_{1},t_{2}]  \\ 
       \gamma(t_{2}-t_{1})  \ , & t > t_{2}  \ . 
           \end{array}\right.
\end{equation}
Then Eq. (\ref{Pave}) becomes,
\begin{eqnarray}
P(x) &=& \int_{t_{1}}^{t_{2}} dT  e^{-\gamma(T-t_{1})} \frac{[\gamma(T-t_{1})]^{x-1}}{(x-1)!} x_{c}^{-1} e^{-T/x_{c}}  \nonumber \\ &+& \int_{t_{2}}^{\infty} dT  e^{-\gamma(t_{2}-t_{1})} \frac{[\gamma(t_{2}-t_{1})]^{x-1}}{(x-1)!} x_{c}^{-1} e^{-T/x_{c}}  \nonumber \\
& = &  \frac{e^{-t_{1}/x_{c}}}{\gamma x_{c}}\Big( \frac{1}{1+1/\gamma x_{c}}\Big)^{x} \Big[ 1-\frac{\Gamma(x, \gamma(t_{2}-t_{1}))}{\Gamma(x)}\Big] \nonumber \\ &+& e^{-\gamma(t_{2}-t_{1})} \frac{[\gamma(t_{2}-t_{1})]^{x-1}}{ \Gamma(x)}e^{-t_{2}/x_{c}} \ .
\end{eqnarray}
In the limit $t_{1} \rightarrow 0$ and with $t_{2} \equiv x_{c}$, the functional form of $P(x)$ has only one parameter, the product $\gamma x_{c} \gg 1$, so that
\begin{eqnarray}
P(x) &=& \frac{1}{\gamma x_{c}}\Big[ 1-\frac{\Gamma(x, \gamma x_{c})}{\Gamma(x)}\Big] \nonumber \\ &+& e^{-(\gamma x_{c}+1)}  \frac{[\gamma x_{c}]^{x-1}}{ \Gamma(x)}
\label{Pxexact}
\end{eqnarray}
In Fig. \ref{figS3} we plot $P(x)$ for several values of the parameter $\gamma x_{c}$, where each curve demonstrates two common features, (i) a uniform distribution of career longevity $x$ for $1 \leq x \lesssim \gamma x_{c}$, and (ii) a sharp peak that is centered around $x =  \gamma x_{c}$ which corresponds to  approximately 10\% of careers which are stellar. Averaging the $P(x)$ over a distribution $P(\gamma)$ of talent values $\gamma$ that is approximately normal, as in the case of the prowess pdfs in Fig. \ref{fig4}, would result in a qualitatively similar $P(x)$ which is peaked around the value $x \approx \overline{\gamma} x_{c}$. The resulting distribution would be essentially ``bimodal'', with one mode corresponding to ``stellar'' careers distributed for $x \approx \overline{\gamma} x_{c}$, and a mode corresponding to less-substantial careers for $x \lesssim \overline{\gamma} x_{c}$, just as in the case of the convex progress rate for $\alpha >1$,  both of which do not agree with the statistical regularity in the empirical data (Fig.~3)   which occurs over several orders of magnitude.

In our model, we assume that termination is due to external factors. A more complex model might include the possibility that termination is due to endogenous factors, e.g. a reduced level of productivity below a predetermined employment threshold at any given time. This type of endogenous termination is more difficult to model, since it correlates the progress $\delta x / \delta t$ with the termination probability $r(T)$, whereas above they are assumed to evolve independently.  We leave this more complex model as an open avenue of research.

\newpage
\clearpage

\begin{widetext}

\begin{table}
\caption{ Summary of data sets for each journal. Total number N of unique (but possibly degenerate) name
identifications.
 $N^{*}$ is the number of unique name identifications after pruning the data set of incomplete careers.}
\begin{tabular}{@{\vrule height 10.5pt depth4pt  width0pt}cc||c||c||c||}\\
\noalign{
\vskip-11pt}
\vrule depth 6pt width 0pt \textbf{\em Journal}  &  Years & Articles & Authors, N & $ N^{*}$\\
\hline  
Nature &  1958-2008 & 65,709 &  130,596  & 94,221 \\
Science &  1958-2008 & 48,169 & 109,519  &  82,181\\
PNAS &  1958-2008 & 84,520 &  182,761  & 118,757\\
PRL &  1958-2008 & 85,316 & 112,660 & 72,102 \\
CELL &  1974-2008 & 11,078 &  31,918  & 23,060 \\
NEJM &  1958-2008 & 17,088 &  66,834  & 49,341 \\
\hline
\end{tabular}
\label{table:journals}
\end{table}

\begin{table}
\caption{ Data summary for the pdfs of career statistical metrics. 
The values $\alpha$ and $x_{c}$ are determined for each career longevity pdf $P(x)$ and each career success pdf $P(z)$
via least-squares method using the functional form given by Eq.~[5]. 
We calculate the Gamma pdf average  $\langle x \rangle$, the standard deviation $\sigma$, and  the extreme threshold
value $x^{*}$ at the $f=0.019$ significance level using the corresponding values of $\alpha$ and $x_{c}$. 
The units for each metric  are indicated in parenthesis alongside the league in the first column.} For publication
distributions, the career longevity metric $x$ is measured in years.
\begin{tabular}{@{\vrule height 10.5pt depth4pt  width0pt}lcc||ccccc} &\multicolumn2l{Least-square
values}&\multicolumn5l{Gamma pdf values}\\
\noalign{
\vskip-11pt} Professional League,\\
\cline{2-8}
\vrule depth 6pt width 0pt (success metric)&  $\alpha$  & $x_{c}$ & $ \langle x \rangle$ & $\sigma$ & $x^{*}$ &$
\frac{x^{*}}{\langle x \rangle}$ &$ \frac{x^{*}}{\sigma}$\\
\hline MLB, (H)  & 0.76 $\pm$ 0.02 & 1240 $\pm$ 150 &  300 &  610 &   2400 &   7.8 &   3.9 \\
MLB, (RBI) &  0.76 $\pm$ 0.02& 570 $\pm$ 80 & 140 &  280 &   1100 &   7.8 &   3.9 \\
NBA, (Pts) &  0.69 $\pm$ 0.02 & 7840 $\pm$ 760 & 2400 &  4400 &   17000 &   7.0 &   3.9 \\
NBA, (Reb) &  0.69 $\pm$ 0.02 &  3500 $\pm$ 130  & 1100 &   2000 &   7600 &   6.9 &   3.9 \\
\hline\\
\end{tabular}
\begin{tabular}{@{\vrule height 10.5pt depth4pt  width0pt}lcc||ccccc} &\multicolumn2l{Least-square
values}&\multicolumn5l{Gamma pdf values}\\
\noalign{
\vskip-11pt} Professional League,\\
\cline{2-8}
\vrule depth 6pt width 0pt (opportunities)&  $\alpha$  & $x_{c}$ & $ \langle x \rangle$ & $\sigma$ & $x^{*}$ &$
\frac{x^{*}}{\langle x \rangle}$ &$ \frac{x^{*}}{\sigma}$\\
\hline
 KBB, (AB) &  0.78 $\pm$ 0.02&  2600 $\pm$ 320 & 580  & 1200 &   4700 &   8.2 &   3.9 \\
 MLB, (AB) &  0.77 $\pm$ 0.02 &  5300 $\pm$ 870  & 1200 &   2500 &   9700 &   8.1 &   3.9 \\
 MLB, (IPO) &  0.72 $\pm$ 0.02 &   3400 $\pm$ 240 &  950 &   1800 &   6900 &   7.3 &   3.9 \\
 KBB, (IPO) &  0.69 $\pm$ 0.02 &   2800 $\pm$ 160 & 840 &   1500 &   5900 &   7.0  & 3.9 \\
 NBA, (Min)  & 0.64 $\pm$ 0.02 &  20600 $\pm$ 1900 & 7700 &   12600 &   48800 &   6.4 &   3.9  \\
 UK, (G) &  0.56 $\pm$ 0.02 & 138 $\pm$ 14 & 61  &   92 &  360 &   5.8 &   3.9 \\
\hline\\
\end{tabular}

\begin{tabular}{@{\vrule height 10.5pt depth4pt  width0pt}lcc||} &\multicolumn2l{Least-square values}\\
\noalign{
\vskip-11pt} Academic Journal,  \\
\cline{2-3}
\vrule depth 6pt width 0pt (career length in years)&  $\alpha$  & $x_{c}$  \\
\hline
 Nature &  0.38 $\pm$ 0.03 & 9.1 $\pm$ 0.2 \\
 PNAS &  0.30 $\pm$ 0.02 &  9.8 $\pm$ 0.2   \\
 Science &  0.40 $\pm$ 0.02 &  8.7 $\pm$ 0.2  \\
 CELL &  0.36 $\pm$ 0.05 &   6.9 $\pm$ 0.2  \\
 NEJM & 0.10 $\pm$ 0.02 & 10.7 $\pm$ 0.2  \\
 PRL &  0.31 $\pm$ 0.04 & 9.8 $\pm$ 0.3  \\
\hline
\end{tabular}
\label{table:stats}
\end{table}

\end{widetext}


\begin{thebibliography}{99}

\bibitem{Matthew}   Merton~RK (1968) The Matthew effect in science. {\it Science} {\bf 159}: 56--63.

\bibitem{AccumAdvProdDiff} Allison~PD, Stewart~JA (1974) Productivity differences among scientists: Evidence for accumulative advantage.
{\it Am. Soc. Rev.} {\bf 39}: 596--606.

\bibitem{DeSollaPrice} De Solla Price~D (1976) A general theory of bibliometric and other cumulative advantage processes. {\it J. Am. Soc. Inf. Sci.} {\bf 27}: 292--306.

\bibitem{CumAdvInequality} Allison~PD, Long~SL, Krauze TK (1982) Cumulative advantage and inequality in science. {\it Am. So. Rev.} {\bf 47}: 615--625.

\bibitem{MatthewIII} Merton~RK (1988) The Matthew effect in science, II: Cumulative advantage and the symbolism of
intellectual property. {\it ISIS} {\bf 79}: 606--623.

\bibitem{MatthewEducation} Walberg~JH, Tsai~S (1983) Matthew effects in education. {\it American Educational Research Journal} {\bf 20}: 359--373.

\bibitem{MatthewReading} Stanovich~KE (1986) Matthew effects in reading: some consequences of individual differences in the acquisition of literacy.
{\it Reading Research Quarterly} {\bf 21}: 360--407.

\bibitem{MatthewCountries} Bonitz~M, Bruckner~E, Scharnhorst~A (1997)  Characteristics and impact of the Matthew effect for countries. {\it Scientometrics} {\bf 40}: 407--422.

\bibitem{MatthewII}   ``For to all those who have, more will be given, and they will have an abundance; but from those
who have nothing, even what they have will be taken away." Matthew 25:29, New Revised Standard Version.


\bibitem{ColeCole} Cole~S, Cole~JR (1967) Scientific output and recognition: A study in the operation of the reward system in science. {\it Am. Soc. Rev.} {\bf 32}: 377--390.

\bibitem{RAFsoccer} Helsen~WF,  Starkes~JL, Van Winckel~J (1998) The influence of relative age on success and dropout in male soccer players. {\it American Journal of Human Biology} {\bf 10}: 791--798.

\bibitem{RAFsport}  Musch~J, Grondin~S (2001) Unequal competition as an impediment to personal development: A review of the relative age effect in sport. {\it Developmental Review} {\bf 21}: 147--167.
  






\bibitem{CareerTrajectory} Simonton~DK (1997) Creative productivity: A predictive 
and explanatory model of career trajectories and landmarks. {\it Psychological Review} {\bf 104}: 66--89.


\bibitem{firing}  Segalla~M,  Jacobs-Belschak~G,  M\"uller~C (2001) 
Cultural influences on employee termination decisions: Firing the Good, Average or the Old? 
{\it European Management Journal} {\bf 19}: 58--72.


\bibitem{LifetimeStats}  Lawless~JF (2003) {\it Statistical models and methods for lifetime data, 2 ed.} (John Wiley \&
Sons, USA).

\bibitem{ReedH}  Reed~WJ,  Hughes~BD (2002) 
From gene families and genera to incomes and internet file sizes: Why power laws are so common in nature. 
{\it Phys. Rev. E} {\bf 66}: 067103.

\bibitem{Hindex} Hirsch~JE (2005) 
An index to quantify an individual's scientific research output. 
{\it Proc. Natl. Acad. Sci. U.S.A.} {\bf 102}: 16569--16572.


\bibitem{physranking}   Radicchi ~F, Fortunato~S, Markines~B, Vespignani~A (2009) Diffusion of scientific credits
and the ranking of scientists. Phys. Rev. E {\bf 80}, 056103.

\bibitem{music}   Davies~JA (2002)
The individual success of musicians, like that of physicists, follows a stretched exponential distribution.
{\it Eur. Phys. J. B} {\bf 27}: 593--595. 

\bibitem{Huber98} Huber~JC (1998)
Inventive Productivity and the Statistics of Exceedances. 
{\it Scientometrics} {\bf 45}: 33--53.

\bibitem{CiteProd} Petersen~AM, Wang~F, Stanley~HE (2010) Methods for measuring the citations and productivity of scientists
  across time and discipline. {\it Phys. Rev. E} {\bf 81}: 036114 .
  
\bibitem{donato}
Castillo~C, Donato ~D \& Gionis~A, {\it Estimating number of citations using author reputation}. Lecture Notes in
Computer Science, (Springer-Verlag, Berlin, 2007).

\bibitem{GenPubAuth} Sidiropoulos~A \& Manolopoulos~Y  (2006)  Generalized comparison of graph-based ranking algorithms
for publications and authors. {\it Journal for Systems \& Software} {\bf 79}: 1679--1700.

\bibitem{Hgen} Sidiropoulos~A, Katsaros~D \& Manolopoulos~Y  (2007)  Generalized Hirsch h-index for disclosing latent facts in
citation networks. {\it Scientometrics} {\bf 72}: 253--280.

\bibitem{hindexResearchers} Batista~PD, Campiteli ~MG, Martinez~AS (2006) Is it possible to compare researchers with
different scientific interests? {\it Scientometrics} {\bf 68}: 179--189.

\bibitem{gappaper} Petersen~AM, Stanley~HE, Succi~S (2011) Statistical regularities in the rank-citation profile of scientists. {\it Submitted.}
ArXiv preprint: (1103.2719 [physics.soc-ph]).

\bibitem{UnivCite} Radicchi~F, Fortunato~S \& Castellano~C (2008) Universality of citation distributions: Toward an
objective measure of scientific impact. {\it Proc. Natl. Acad. Sci. USA }{\bf 105}: 17268--17272.

\bibitem{BB}  Petersen~AM,   Jung~W-S \& Stanley ~HE (2008)
On the distribution of career longevity and the evolution of home-run prowess in professional baseball. 
{\it Europhysics Letters} {\bf 83}, 50010.

\bibitem{Tennis} Radicchi~F (2011) 
Who is the best player ever? A complex network analysis of the history of professional tennis. {\it PLoS ONE} {\bf 6}: e17249.

\bibitem{SportsStats} 
Sean Lahman's Baseball Archive:\\
http://baseball1.com/index.php \\
Korean Professional Baseball League: http://www.inning.co.kr \\
Data Base Sports Basketball Archive:\\
http://www.databasebasketball.com/ \\
Barclays Premier League: http://www.premierleague.com/

\bibitem{journallimiations} ISI Web of Knowledge: www.isiknowledge.com/. 





\bibitem{110PhsRev} Redner ~S (2005) Citation statistics from 110 years of Physical Review. {\it Physics Today} {\bf
58}: 49-54.

\bibitem{BB3}  Petersen~AM, Penner~O, Stanley~HE. (2011) 
Methods for detrending success metrics to account for inflationary and deflationary factors.
{\it Eur. Phys. J. B} {\bf 79}, 67--78. Preprint title: Detrending career statistics in professional Baseball:
accounting for the Steroids Era and beyond. e-print arXiv:1003.0134.

\bibitem{CompSocialScience}  Lazer~D, {\it et al} (2009)
Computational social science. 
{\it Science} {\bf 323}: 721--723.

\bibitem{SocialDynamics}   Castellano~C,   Fortunato~S \&   Loreto~V (2009) 
 Statistical physics of social dynamics. 
 {\it Rev. Mod. Phys.} {\bf 81}: 591--646.

\bibitem{socpowlaw}   Liljeros~F,  {\it et al} (2001)
The web of human sexual contacts. 
{\it Nature} {\bf 411}: 907--908 .

\bibitem{PAsex}   de Blasio~BF,   Svensson~A, \&   Liljeros~F (2007)
 Preferential attachment in sexual networks. 
 {\it Proc. Natl. Acad. Sci. U.S.A.} {\bf 104}: 10762--10767.



\bibitem{MEJN}  Newman~MEJ (2005)
Power laws, Pareto distributions and Zipf's law. 
{\it Contemporary Physics} {\bf 46}: 323--351.

\bibitem{econpowlaw} Farmer~JD,   Shubik~M \&  Smith~E (2005)
  Is economics the next physical science? 
 {\it Physics Today} {\bf 58(9)}: 37--42.

\bibitem{Barabasibursts}  Barab\'asi~AL (2005)
 The origin of bursts and heavy tails in human dynamics. 
{\it Nature} {\bf 435}: 207--211.

\bibitem{wattsSW}  Watts~DJ,   Strogatz~SH (1998)
 Collective dynamics of small-world networks. 
{\it Nature} {\bf 393}: 440--442.


\bibitem{mobility}  Gonz\'alez~MC,  Hidalgo~CA \&  Barab\'asi~AL (2008)
Understanding individual human mobility patterns.
{\it Nature} {\bf 453}: 779--782.

\bibitem{Emails} Malmgren~RD, {\it  et al} (2008)
A Poissonian explanation for heavy tails in e-mail communication. 
{\it Proc. Natl. Acad. Sci. } {\bf 105}: 18153--18158.

\bibitem{SornettePNAS} Crane~R, Sornette~D (2008) Robust dynamic classes revealed by measuring the response function of
a social system. 
{\it Proc. Natl. Acad. Sci. }  105: 15649--15653.

\bibitem{linking}  Leskovec~J,  {\it  et al} (2008)
 Microscopic evolution of social networks. 
 {\it Proceeding of the ACM SIGKDD International Conference on Knowledge Discovery and Data Mining,  Association for
Computing Machinery}: 462--470.


\bibitem{FOMC} Petersen~AM, Wang~F, Havlin~S, Stanley~HE (2010) Quantitative law describing market dynamics before and after interest rate change. {\it Phys. Rev. E} {\bf 81}: 066121.

\bibitem{GravityHighway} Jung~W-S,  Wang~F,  Stanley~HE (2008) Gravity model in the Korean highway. {\it Europhys. Lett.} {\bf 81}: 48005.

\bibitem{MetroHopper1} Lee~K, Jung W-S, Park~JS, Choi~MY (2008) Statistical analysis of the
Metropolitan Seoul Subway System: Network structure and passenger flows.
{\it Physica A} {\bf 387}: 6231--6234


\bibitem{HumanMobility2} Song~C, Koren~T, Wang~P, Barab\'asi~AL (2010) Modeling the scaling properties of human mobility. {\it Nature Physics} {\bf 6}: 818--823.

\bibitem{firstpassage}   Redner~S, {\it A guide to first-passage processes}. (Cambridge University Press, UK, 2001).

\bibitem{spatialPoisson}    Larson~RC,  Odoni~AR, {\it Urban operations research}. 2nd Ed. (Dynamic Ideas, USA,
2007).







\end{thebibliography}

\begin{thebibliography}{99}



\bibitem[S1]{Huber98s}  Huber~JC (1998) 
Inventive Productivity and the Statistics of Exceedances.
 {\it Scientometrics} {\bf 45}: 33.

\bibitem[S2]{physranking2}  Radicchi ~F, Fortunato~S, Markines~B, Vespignani~A (2009) Diffusion of scientific credits and
the ranking of scientists. Phys. Rev. E {\bf 80}, 056103.

\bibitem[S3]{BBs}  Petersen~AM,   Jung~W-S \& Stanley ~HE (2008) On the distribution of career longevity and the evolution of
home-run prowess in professional baseball. 
{\it Europhysics Letters} {\bf 83}, 50010.

\bibitem[S4]{2randv}  Glen~A,  Leemis~L  \&   Drew~J (2004)
 Computing the distribution of the product of two continuous random variables. 
{\it Computational Stat. \& Data Analysis} {\bf 44}, 451.

\bibitem[S5]{BB32}  Petersen~AM, Penner~O, Stanley~HE (2011)   
Methods for detrending success metrics to account for inflationary and deflationary factors.
{\it Eur. Phys. J. B} {\bf 79}, 67--78. Preprint title: Detrending career statistics in professional Baseball:
accounting for the Steroids Era and beyond. e-print arXiv:1003.0134.

\bibitem[S6]{CareerTrajectorys} Simonton~DK (1997) Creative productivity: A predictive 
and explanatory model of career trajectories and landmarks. {\it Psychological Review} {\bf 104}: 66-89.

\end{thebibliography}
\end{document}